\newif\ifdoubleblind
\newif\ifacm

\ifacm
	\documentclass[sigconf]{acmart}
\else
	\documentclass[conference]{IEEEtran}
\fi

\usepackage{amsmath}
\DeclareMathOperator*{\argmax}{arg\,max}

\usepackage{multirow}
\usepackage{tabularx,booktabs}
\usepackage{xspace}

\usepackage[binary-units=true]{siunitx}
\usepackage{tikz}
\usetikzlibrary{positioning}

\ifacm
	\usepackage[nolist]{acronym}
\usepackage{booktabs} 
\usepackage{subfig}
\usepackage{balance}

\usepackage{array}
\usepackage{graphicx}
\usepackage{wasysym}
\usepackage{xparse}

\usepackage{makecell}
\newcolumntype{Y}{>{\centering\arraybackslash}X}

\hypersetup{draft}

\renewcommand\footnotetextcopyrightpermission[1]{} 
\else
	\usepackage{graphicx}
\usepackage{epstopdf}
\usepackage{standalone}
\usepackage[nolist ]{acronym}
\usepackage{tcolorbox}
\usepackage{pdfpages}
\usepackage[bookmarks=false]{hyperref}
\usepackage{pdfcomment}
\usepackage{hologo}

\usepackage{xstring}

\usepackage[utf8]{inputenc}
\usepackage{marvosym}

\usepackage{algorithm}
\usepackage{algpseudocode}

\usepackage{psfrag}
\usepackage{graphicx}

\usepackage{rotating}

\renewcommand{\baselinestretch}{1}

\usepackage[caption=false,font=footnotesize]{subfig}
\usepackage{stfloats}
\fi
\renewcommand{\baselinestretch}{0.995}

\begin{document}

\newcommand{\paperTitle}{PARRoT: Predictive Ad-hoc Routing Fueled by Reinforcement Learning and Trajectory Knowledge}
\newcommand{\paperAuthors}{Benjamin Sliwa, Cedrik Schüler, Manuel Patchou, and Christian Wietfeld}
\newcommand{\paperEmails}{$\{$Benjamin.Sliwa, Cedrik.Schueler, Manuel.Mbankeu, Christian.Wietfeld$\}$@tu-dortmund.de}

\newcommand\single{1\textwidth}
\newcommand\double{.48\textwidth}
\newcommand\triple{.32\textwidth}
\newcommand\quarter{.24\textwidth}
\newcommand\singleC{1\columnwidth}
\newcommand\doubleC{.475\columnwidth}

\newcommand{\figurePadding}{0pt}
\newcommand{\figureTopPadding}{\figurePadding}
\newcommand{\figureBottomPadding}{\figurePadding}
\newcommand{\red}[1]{\colorbox{red}{\textbf{TODO}: #1}}
\newcommand{\batmobile}{B.A.T.Mobile\xspace}
\newcommand{\protocol}{\ac{PARRoT}\xspace}
\newcommand{\parrot}{\ac{PARRoT}\xspace}
\newcommand{\reward}{$V_f$\xspace}

\newcommand\tikzFig[2]
{
	\begin{tikzpicture}
		\node[draw,minimum height=#2,minimum width=\columnwidth,text width=\columnwidth,pos=0.5]{\LARGE #1};
	\end{tikzpicture}
}

\newcommand{\dummy}[3]
{
	\begin{figure}[b!]  
		\begin{tikzpicture}
		\node[draw,minimum height=6cm,minimum width=\columnwidth,text width=\columnwidth,pos=0.5]{\LARGE #1};
		\end{tikzpicture}
		\caption{#2}
		\label{#3}
	\end{figure}
}

\newcommand\pos{h!tb}

\newcommand{\basicFig}[7]
{
	\begin{figure}[#1]  	
		\vspace{#6}
		\centering		  
		\includegraphics[width=#7\columnwidth]{#2}
		\caption{#3}
		\label{#4}
		\vspace{#5}	
	\end{figure}
}
\newcommand{\fig}[4]{\basicFig{#1}{#2}{#3}{#4}{0cm}{0cm}{1}}

\newcommand\sFig[2]{\begin{subfigure}{#2}\includegraphics[width=\textwidth]{#1}\caption{}\end{subfigure}}
\newcommand\vs{\vspace{-0.3cm}}
\newcommand\vsF{\vspace{-0.4cm}}

\newcommand{\subfig}[3]
{%
	\subfloat[#3]%
	{%
		\includegraphics[width=#2\textwidth]{#1}%
	}%
	\hfill%
}

\newcommand\circled[1] 
{
	\tikz[baseline=(char.base)]
	{
		\node[shape=circle,draw,inner sep=1pt] (char) {#1};
	}\xspace
}
\begin{acronym}
	\acro{B.A.T.M.A.N.}{Better Approach To Mobile Ad-hoc Networking}
	\acro{OMNeT++}{Objective Modular Testbed in C++}
	\acro{LIMITS}{LIghtweight Machine learning for IoT Systems}
	\acro{LIMoSim}{Lightweight ICT-centric Mobility Simulation}
	\acro{MANET}{Mobile Ad-hoc Network}
	\acro{VANET}{Vehicular Ad-hoc Network}
	\acro{FANET}{Flying Ad-hoc Network}
	\acro{UAV}{Unmanned Aerial Vehicle}
	\acro{ITS}{Intelligent Transportation System}
	\acro{AODV}{Ad-hoc On-demand Distance Vector}
	\acro{DYMO}{Dynamic MANET On Demand Routing Protocol}
	\acro{OLSR}{Optimized Link State Routing}
	\acro{GPSR}{Greedy Perimeter Stateless Routing in Wireless Networks}
	\acro{DSDV}{Destination-Sequenced Distance Vector}
	\acro{PARRoT}{Predictive Ad-hoc Routing fueled by Reinforcement learning and Trajectory knowledge}
	\acro{PDR}{Packet Delivery Ratio}
	\acro{UDP}{User Datagram Protocol}
	\acro{CBR}{Constant Bitrate}
	\acro{LOS}{Line-of-Sight}
	\acro{RREQ}{Route Request}
	\acro{RREP}{Route Reply}
	\acro{LET}{Link Expiry Time}
	\acro{LIMoSim}{LIghtweight ICT-centric Mobility Simulation}
	\acro{MCN}{Multi-hop Cellular Network}
	\acro{MPR}{Multipoint Relay}
	\acro{CRS-MP}{Centralized Routing Scheme with Mobility Prediction}
	\acro{SDN}{Software-defined Networking}
	\acro{ANN}{Artificial Neural Network}
	\acro{DDD}{Distributed Dispersion Detection}
	\acro{TTL}{Time to Live}
	\acro{SEQ}{Sequence Number}
	\acro{HWMP}{Hybrid Wireless Mesh Protocol}
	\acro{MAC}{Medium Access Control}
	\acro{GPS}{Global Positioning System}
\end{acronym}

\title{\paperTitle}

\ifacm
\else

	\title{\paperTitle}

	\ifdoubleblind
	\author{\IEEEauthorblockN{\textbf{Anonymous Authors}}
		\IEEEauthorblockA{Anonymous Institutions\\
			e-mail: Anonymous Emails}}
	\else
	\author{\IEEEauthorblockN{\textbf{\paperAuthors}}
		\IEEEauthorblockA{Communication Networks Institute,	TU Dortmund University, 44227 Dortmund, Germany\\
			e-mail: \paperEmails}}
	\fi
	
	\maketitle

\fi




\begin{abstract}
	
%
%
Swarms of collaborating \acp{UAV} that utilize ad-hoc networking technologies for coordinating their actions offer the potential to catalyze emerging research fields such as autonomous exploration of disaster areas, demand-driven network provisioning, and near field packet delivery in \acp{ITS}.
%
%
As these mobile robotic networks are characterized by high grades of relative mobility, existing routing protocols often fail to adopt their decision making to the implied network topology dynamics.
%
%
For addressing these challenges, we present \protocol as a novel machine learning-enabled routing protocol which exploits mobility control information for integrating knowledge about the future motion of the mobile agents into the routing process.
%
%
The performance of the proposed routing approach is evaluated using comprehensive network simulation. In comparison to established routing protocols, \protocol achieves a massively higher robustness and a significantly lower end-to-end latency.

\end{abstract}

\ifacm

\fi

\maketitle
\begin{tikzpicture}[remember picture, overlay]
\node[below=5mm of current page.north, text width=20cm,font=\sffamily\footnotesize,align=center] {Accepted for presentation in: 2021 IEEE 93rd Vehicular Technology Conference (VTC-Spring)\vspace{0.3cm}\\\pdfcomment[color=yellow,icon=Note]{
@InProceedings\{Sliwa/etal/2021a,\\
  author    = \{Benjamin Sliwa and Cedrik Schüler and Manuel Patchou and Christian Wietfeld\},\\
  booktitle = \{2021 IEEE 93rd Vehicular Technology Conference (VTC-Spring)\},\\
  title     = \{\{PARRoT\}: \{P\}redictive ad-hoc routing fueled by reinforcement learning and trajectory knowledge\},\\
  year      = \{2021\},\\
  address   = \{Helsinki, Finland\},\\
  month     = \{Apr\},\\
\}
}};
\node[above=5mm of current page.south, text width=15cm,font=\sffamily\footnotesize] {2021~IEEE. Personal use of this material is permitted. Permission from IEEE must be obtained for all other uses, including reprinting/republishing this material for advertising or promotional purposes, collecting new collected works for resale or redistribution to servers or lists, or reuse of any copyrighted component of this work in other works.};
\end{tikzpicture}

\section{Introduction}

\begin{figure}[]  	
	\vspace{0cm}
	\centering		  
	\includegraphics[width=1.0\columnwidth]{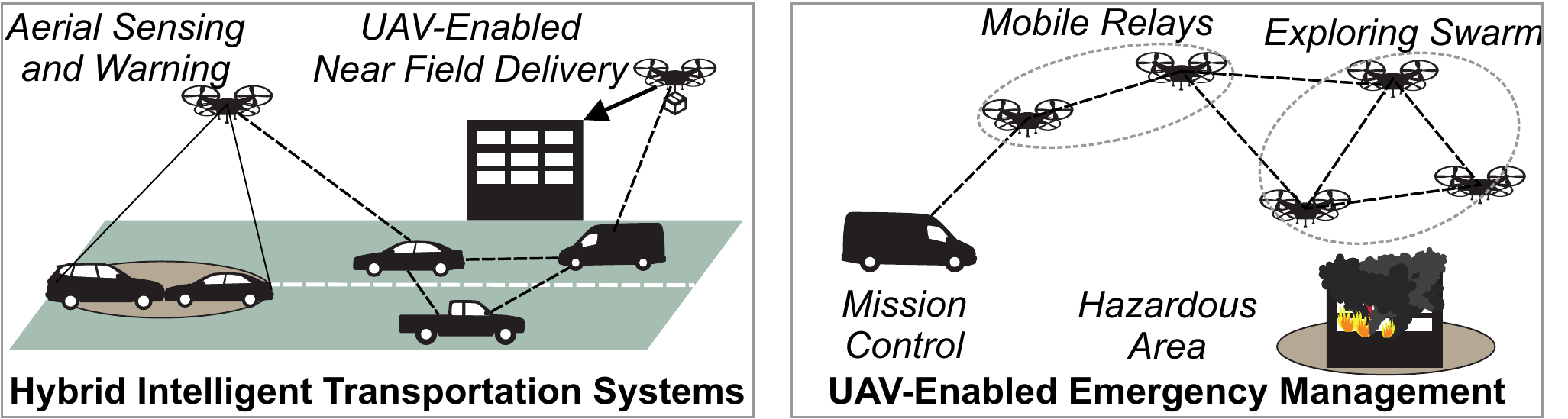}
	\vspace{-0.5cm}	
	\caption{Example applications of \ac{UAV}-enabled wireless mesh networks.}
	\label{fig:scenario}
	\vspace{-0.5cm}	
\end{figure}

%
%
Collaborating autonomous drones that coordinate their actions using \acp{FANET} offer the potential to efficiently perform important disaster relief tasks --- e.g., remote sensing and network provisioning --- without risking the lives of human helpers \cite{Zeng/etal/2019a}.
%
%
A closely related emerging research field is the integration of small-scale \acp{UAV} into future \acp{ITS} \cite{Menouar/etal/2017a} for applications such as aerial traffic monitoring \cite{Sliwa/etal/2019b} and \ac{UAV}-aided near field delivery \cite{Patchou/etal/2019a}. While the latter concept has been initially proposed for reducing the delivery time in inner cities, its inherent avoidance of direct human-to-human interaction also makes it a promising candidate for increasing the delivery safety during the COVID-19 pandemic \cite{Chamola/etal/2020a}.
An illustration about different applications of \ac{UAV}-based \acp{FANET} is shown in  Fig.~\ref{fig:scenario}.
%
%
For enabling these novel use-cases, the provision of efficient and reliable means of communication even in challenging environments is an important prerequisite. However, established \ac{MANET} routing protocols can often barely cope with the small channel coherence time and the network topology dynamic implied by the high grade of relative mobility.
%
%
Anticipatory mobile networking \cite{Bui/etal/2017a} has been proposed for explicitly addressing the interdependency of mobility and communication by integrating \emph{context} knowledge into the corresponding decision processes. This novel communications paradigm has a strong relationship to the usage of machine learning for optimizing wireless communication networks \cite{Wang/etal/2020a} which manifests in the trend of replacing complex mathematical models by learned representations of the corresponding phenomena \cite{Ali/etal/2020a}. Moreover, it has catalyzed the emergence of novel performance evaluation methods that are capable of replacing computationally expensive entity-based modeling with machine learning-based end-to-end models \cite{Sliwa/Wietfeld/2019c}.
%
%
In this paper, we present \protocol as a novel reinforcement learning-enabled cross-layer routing protocol which leverages application layer knowledge from the mobility control routines for proactively optimizing the robustness of vehicular routing paths.
%
%
The remainder of the paper is structured as follows. After discussing the related work in Sec.~\ref{sec:related_work}, the novel \protocol protocol is presented in Sec.~\ref{sec:approach}. Afterwards, an overview about the methodological aspects of the simulative performance evaluation is given in Sec.~\ref{sec:methods}. Finally, detailed simulation results are provided and discussed in Sec.~\ref{sec:results}.
%
%
%

%
%

\section{Related Work} \label{sec:related_work}

%
%
A wide range of solution approaches for specific applications and different kinds of vehicular networks has been proposed by literature. 
%
%
Comprehensive summaries about existing protocols for highly mobile networks are presented by the authors of \cite{Oubbati/etal/2019a} and \cite{Nazib/Moh/2020a}. Moreover, Cavalcanti et al. \cite{Cavalcanti/etal/2018a} provide an empirical analysis of the popularity of existing routing protocols and performance evaluation methods in the context of vehicular networking.
%
%
Classically, \ac{MANET} routing protocols have been classified into \emph{reactive} --- e.g., \ac{AODV} and \ac{DYMO} --- and \emph{proactive} --- e.g., \ac{DSDV}, \ac{OLSR}, and \ac{B.A.T.M.A.N.} --- methods.
%
%
However, the need to pay attention to the interdependency between mobility and communication has lead to the rise of \emph{geo-based} routing approaches such as \ac{GPSR} which integrate position and velocity information into their decision making.
%
%
In extension, \emph{geo-predictive} approaches such as \batmobile consider the anticipated future relative mobility of the vehicles within the routing process \cite{Sliwa/etal/2016b}.
%
%
Similarly, Song et al. \cite{Song/etal/2018a} present an extended variant of \ac{OLSR} which uses Kalman filter-based mobility prediction for optimizing the \ac{MPR} determination process.

%
%
\emph{Reinforcement learning} can be regarded as a step towards zero touch optimization of wireless communication systems. Hereby, an \emph{agent} learns to autonomously perform favorable \emph{actions} in a defined  \emph{environment} through observation of the \emph{rewards} of previously taken actions. Machine learning-enabled routing methods have been proposed by different researchers.
%
%
The authors of \cite{Tang/etal/2019a} utilize an \ac{ANN}-enabled centralized routing approach using \ac{SDN} for \acp{VANET} delay minimization. Similar to our work, their proposed routing method \ac{CRS-MP} takes into account mobility predictions of the mobile vehicles. 
%
%
Oddi et al. \cite{Oddi/etal/2012a} use geo-based routing metrics jointly with Q-Learning-enabled reinforcement learning. Similar to \batmobile, and as further discussed in Sec.~\ref{sec:approach}, this method represents important groundwork for the novel \protocol protocol.
%
%
Due to the availability of data analysis tools such as \ac{LIMITS} \cite{Sliwa/etal/2020c} which allow to automatically derive \texttt{C++} implementations of trained prediction models, it can be expected that pervasive machine learning will be one of the key enablers for future network generation. 
%
%
%

%
%
Another ongoing development is the partial convergence of cellular and ad-hoc networking paradigms, illustrated by a growing interest in integrating single- and multi-hop device-to-device communication into cellular networks and manifested by concepts such as \acp{MCN} \cite{Coll-Perales/etal/2019a}.
As a consequence, novel developments in the ad-hoc networking domain might also have an impact on future cellular network generations \cite{Ali/etal/2020a}.
\section{Machine Learning-Enabled Wireless Mesh Routing with \protocol} \label{sec:approach}

%
%
\begin{figure}[]  	
	\vspace{0cm}
	\centering		  
	\includegraphics[width=1.0\columnwidth]{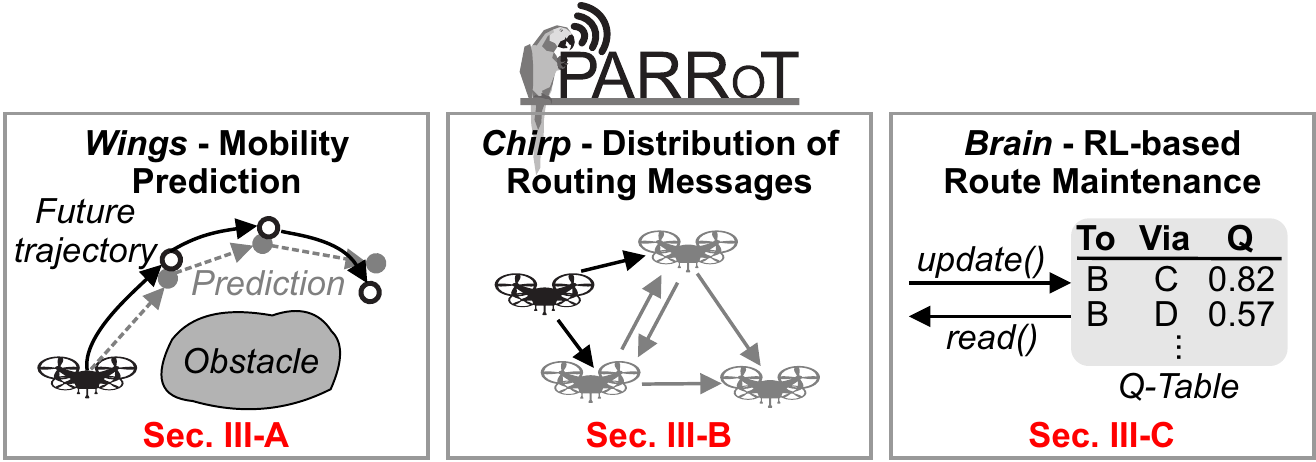}
	\vspace{-0.5cm}	
	\caption{Overall system architecture model of the \protocol routing protocol.}
	\label{fig:architecture}
	\vspace{-0.5cm}	
\end{figure}
The overall system architecture model of the proposed \protocol is shown in Fig.~\ref{fig:architecture}. \protocol consists of three logical core components which are further explained in the following paragraphs.

\subsection{\protocol Wings - Cross-Layer Mobility Prediction} \label{sec:mobility}

\protocol utilizes a cross-layer approach that leverages knowledge from the mobility control domain for anticipating the relative mobility between the different agents for a defined prediction horizon $\tau$.
For this purpose, each agent estimates its own future position $\mathbf{\tilde{p}}(t+\tau)$ based on the current position $\mathbf{p}(t)$. The result is then propagated to the other nodes of the network via the \emph{chirp} distribution process (see Sec.~\ref{sec:routing}).
If the \protocol agent owns information about the planned trajectory as a sequence of $k$ waypoints, an iterative prediction method consisting of $N = \lfloor \tau / \Delta t \rfloor$ steps is applied whereas $\Delta t$ represents the mobility update interval and the final result is given by $\tilde{\mathbf{p}}(t+\tau) = \tilde{\mathbf{p}}_{N}$. In each iteration $i$, the agent virtually moves towards the current waypoint $\mathbf{w}_{k}$ as

%
%
\begin{equation} \label{eq:mobility_prediction} 
	\tilde{\mathbf{p}}_{i+1} = \tilde{\mathbf{p}}_{i} + \frac{\mathbf{w}_{k}-\tilde{\mathbf{p}}_{i}}{||\mathbf{w}_{k}-\tilde{\mathbf{p}}_{i}||} \cdot v \cdot \Delta t
\end{equation}
%
%
with $\tilde{\mathbf{p}}_0 = \mathbf{p}(t)$ and $v$ being the current velocity. After each iteration, it is checked if the vehicle is now within the radius $r_w$ of the waypoint sphere: If  $||\mathbf{w}_{k}-\tilde{\mathbf{p}}_{i+1}|| \leq r_w$ is fulfilled, the waypoint is considered \emph{reached} and $k$ is incremented.
%
%
%

%
%
As a fallback mechanism for cases where no waypoint information is available, the average slope of the previous $h$ positions is computed and extrapolated for $\tau$. This method is non-iterative and allows to immediately compute the final result as
%
%
\begin{equation}
	\tilde{\mathbf{p}}(t+\tau) = \mathbf{p}(t) + \frac{\tau}{h-1} \sum_{i=1}^{h-1} \frac{\mathbf{p}_i - \mathbf{p_{i-1}}}{\Delta t} .
\end{equation}

\subsection{\protocol Chirp - Dissemination of Routing Messages and Context Knowledge} \label{sec:routing}

\protocol relies on an exchange of \ac{UDP}-based routing messages --- which are referred to as \emph{chirps} in the following --- for adopting the local routing knowledge to the highly dynamic network topology conditions.
%
%
\begin{figure}[]  	
	\vspace{0cm}
	\centering		  
	\includegraphics[width=0.6\columnwidth]{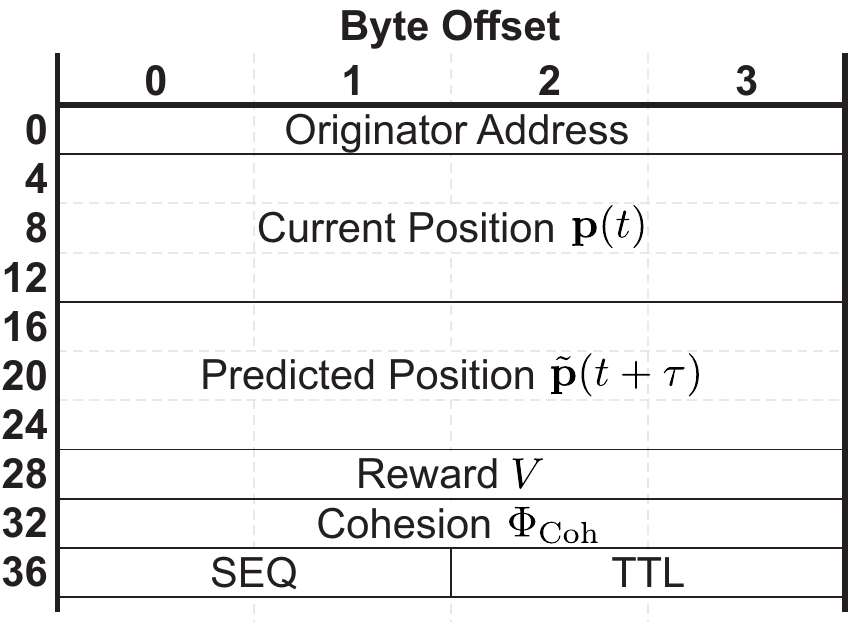}
	\vspace{-0.3cm}	
	\caption{Structure of the chirp data packet with a total size of 40~Byte.}
	\label{fig:packet}
	\vspace{-0.5cm}	
\end{figure}
The structure of the 40~Byte chirp packet is illustrated in Fig.~\ref{fig:packet}.

%
%
\textbf{Chirp message distribution}:
%
%
Each \protocol node periodically generates \emph{chirp} messages based on a fixed interval $\Delta t_{\text{chirp}}$ which are propagated through the network. Sequence numbering is used in order to allow the assessment of the data freshness.
%
%
Upon reception of a \protocol chirp message by node $i$ via forwarder $j$ and originated from $d$, the following steps are performed:
\begin{itemize}
	\item If the \ac{SEQ} for $d$ is not newer than the one of the last received chirp from $d$, the message is discarded and the following steps are omitted.
	\item Otherwise, the received information is utilized to update the local knowledge maintained in the Q-Table based with the extracted values of $V$ and $\Phi_{\text{Coh}}$. The corresponding process is further described in Sec.~\ref{sec:rl}.
	\item After the local handling, the chirp message is forwarded. For this purpose, the contained mobility information $\mathbf{p}(t)$ and $\tilde{\mathbf{p}}(t+\tau)$ of the forwarder $j$ is replaced by the corresponding values of node $i$. The \ac{TTL} is reduced by $1$ and $V$ as well as $\Phi_{\text{Coh}}$ are updated according to Sec.~\ref{sec:rl}.
\end{itemize}
%
%
%

%
%
\begin{figure}[]  	
	\vspace{0cm}
	\centering		  
	\includegraphics[width=0.9\columnwidth]{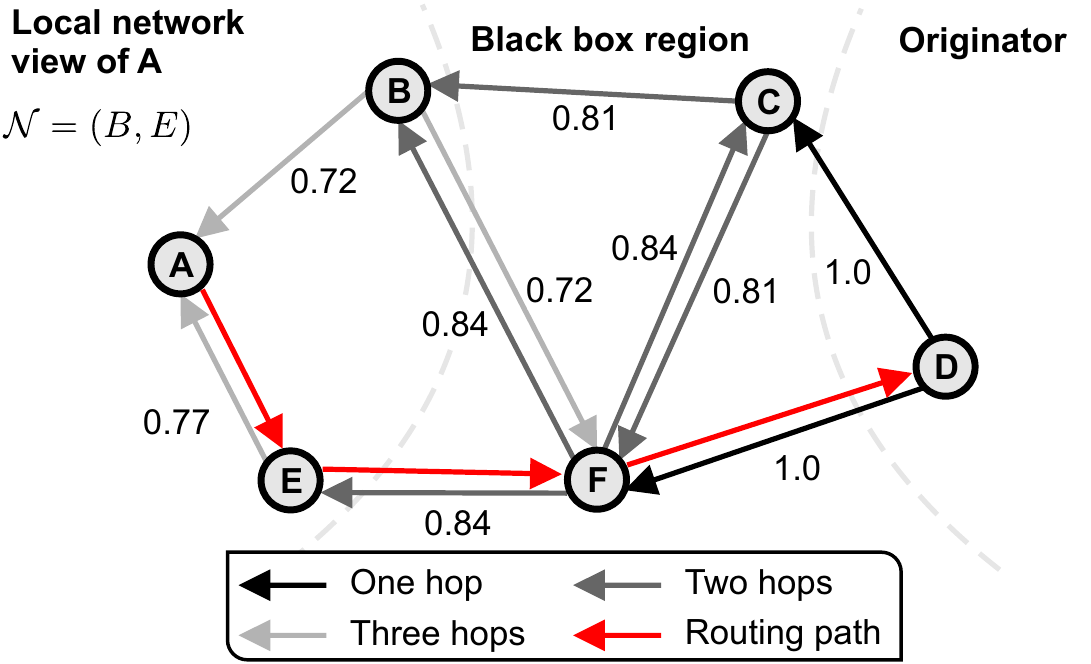}
	\caption{Example for the propagation of chirp messages from originator $D$ and its usage for reverse path routing from $A$ to $D$.}
	\label{fig:routing_example}
	\vspace{-0.5cm}	
\end{figure}
An example for the propagation of chirp messages from node $D$ and its usage for routing data packets from $A$ to $D$ is shown in Fig.~\ref{fig:routing_example}.
\begin{itemize}
	%
	%
	\item Node $D$ generates a chirp and initializes the reward $V_D$ with the highest possible value $1.0$.
	%
	%
	\item The message is received by the one hop neighbor nodes $C$ and $F$ which update their Q-Table entries according to Eq.~\ref{eq:qtable_update}. Both then forward the chirp message with an updated reward which is calculated as $V_{C/F} = Q(D, D)$.
	%
	%
	\item The forwarded messages are received and handled by $B$ and $F$. The rewards are updated as $V_{B/F} = Q(D,\argmax_j Q(D,j))$.
	$C$ and $F$ also mutually receive the chirp message but discard it as the sequence number is outdated.
	%
	%
	\item $A$ finally receives the message originated from $D$ via $B$ and $E$ and updates the corresponding table entries.
	%
	%
	\item For routing messages from $A$ to $D$, node $A$ only knows about its direct neighbors $\mathcal{N}=(B,E)$ and about the existence of the destination node which is hidden beyond a black box network region. In each forwarding step, the message is propagated to neighbor $j = \argmax_j Q(D,j)$.
\end{itemize}
\subsection{\protocol Brain - Reinforcement Learning-Based Route Maintenance} \label{sec:rl}

The routing process of \protocol is inspired by existing approaches and distills initial ideas by the authors of \cite{Oddi/etal/2012a} and \cite{Sliwa/etal/2016b}. It consists of two core components which are explained in the following.

%
%
\textbf{Online routing process}:
%
%
Similar to decentralized approaches such as \ac{B.A.T.M.A.N.} and in contrast to path planning-based protocols such as \ac{OLSR}, each node has only a partial view on the overall network topology. In order to route messages to given destination $d$, each node $i$ only assesses the suitability of its direct neighbors $\mathcal{N}$ for reaching $d$ --- the intermediate network is treated as a black box.
%
%
For this purpose, the numeric values for the end-to-end link quality $Q(d,j)$ are maintained in a \emph{Q-Table}. The online routing decision process can then be formulated as $\argmax_j Q(d,j)$ whereas the one-hop neighbor $j$ with the highest $Q$ value is chosen as a message forwarder. 
The implied maintenance of multiple paths for each destination inherently provides the protocol with self healing capabilities that allow \protocol to quickly recover after failure of nodes.
%
%
%

%
%
\textbf{Update procedure}: 
Upon reception of a chirp message, the receiver node updates its local knowledge about the reverse path to the originator using a modified \emph{Q-Learning} method as
%
%
\begin{equation} \label{eq:qtable_update}
	Q(d, j) = Q(d, j) + \alpha \left[ \gamma(j) \cdot V_j - Q(d, j) \right] 
\end{equation}
whereas $\alpha$ is the learning rate and $V_j$ represents the received reverse path score to the originator $d$ via forwarder $j$ which is extracted from the chirp message. 
%
%
The variable \emph{discount factor} $\gamma(j)$ serves as a \emph{multidimensional routing metric} and is computed as
%
%
\begin{equation} \label{eq:gamma}
	\gamma(j) = \gamma_0 \cdot \Phi_{\text{LET}}(i,j) \cdot  \Phi_{\text{Coh}}(j)
\end{equation}
%
%
with $\gamma_0$ being a constant value for ensuring loop-free routing through a guaranteed metric degradation per hop.

%
%
$\Phi_{\text{LET}}(i,j)$ utilizes an estimation of the \ac{LET} between $i$ and $j$ which takes into the account the results of the mobility prediction process (see Sec.~\ref{sec:mobility}). For a given prediction horizon $\tau$, the metric is computed as
%
%
\begin{equation}
	\Phi_{\text{LET}}(i, j) =
	\begin{cases}
	\sqrt{\frac{\text{LET}(i, j)}{\tau}} &: \text{LET}(i,j) < \tau \\
	1.0 &: \text{else}
	\end{cases}.
\end{equation}
With $\mathbf{\Delta \mathbf{p} = \mathbf{p}_j - \mathbf{p}_i}$ being the relative position and  $\mathbf{\Delta \mathbf{v} = \mathbf{v}_j - \mathbf{v}_i}$ being the relative velocity, the relative trajectory can be written as $\Delta \tilde{\mathbf{p}} = \Delta \mathbf{p} + t \cdot \Delta \mathbf{v}$. 
The $\text{LET}(i, j)$ between nodes $i$ and $j$ represents the time $t$ where the distance between $i$ and $j$ exceeds the maximum communication radius $r_{\text{TX}}$. Thus, $r_{\text{TX}} = \sqrt{\Delta \tilde{\mathbf{p}}_x^2 + \Delta \tilde{\mathbf{p}}_y^2 + \Delta \tilde{\mathbf{p}}_z^2}$ needs to be solved for $t$ in order to determine $\text{LET}(i,j)$. Substituting 
%
%
\begin{align}
	a = \Delta \mathbf{v}_x^2 + \Delta \mathbf{v}_y^2 + \Delta \mathbf{v}_z^2 \\	
	b = 2 \left( \Delta\mathbf{p}_x \Delta\mathbf{v}_x + \Delta\mathbf{p}_y \Delta\mathbf{v}_y + \Delta\mathbf{p}_z \Delta\mathbf{v}_z \right)\\
	c = \Delta \mathbf{p}_x^2 + \Delta \mathbf{p}_y^2 + \Delta \mathbf{p}_z^2 - r_{\text{TX}}^2
\end{align}
allows to derive
%
%
\begin{equation}
	t_{1,2} = \frac{-b \pm \sqrt{b2-4ac}}{2a} .
\end{equation}
Due to the square root, three different cases can be distinguished:
\begin{equation}
	\text{LET}(i,j) =
	\begin{cases}
	0 &: t_1\leq 0 \wedge t_2\leq 0 \\
	t_2 &: t_1\leq 0 \wedge t_2> 0 \\
	0 &: t_1> 0 \wedge t_2> 0 \\
	\end{cases}
\end{equation}
These conditions can be interpreted as follows. In the first case, the link is  and will stay unavailable. In the second case, the link is currently available and will expire at $t_2$. In the last case, the link is expected to become available at $t_1$ and will expire at $t_2$.
%
%
%

%
%
$\Phi_{\text{Coh}}(j)$ is a measure for the neighbor set \emph{coherence} of message forwarder $j$ based on the difference between the neighbor sets $\mathcal{N}$ at the times $t$ and $t-t_h$. According to \cite{Oddi/etal/2012a}, it is derived as
%
%
\begin{equation}
	\Phi_{\text{Coh}}(j) = \sqrt{1-\frac{\mathcal{N}(t) \triangle \mathcal{N}(t-\Delta t)}{\mathcal{N}(t)\cup\mathcal{N}(t-\Delta t)}}
\end{equation}
with operator $\triangle$ being the symmetrical difference between the two considered node sets.

\section{Methodology of the Simulative Performance Evaluation} \label{sec:methods}

In this section, the methodological aspects of the performance evaluation are presented.
%
%
All simulations are performed using the \ac{OMNeT++} framework \cite{Varga/Hornig/2008a} jointly with \texttt{INETMANET} which provides implementation of various \ac{MANET} protocols.
%
%
For the performance analysis, the \ac{PDR} and the end-to-end latency of a video streaming application modeled as \ac{UDP} \ac{CBR} is considered. Within each evaluation run, sender and receiver are randomly chosen from the total set of vehicles.
%
%
The source code of the developed \ac{OMNeT++} implementation for \protocol is provided in an open source manner\footnote{Source code available at \url{https://github.com/cedrikschueler/PARRoT}}.
%
%
%

%
%
\begin{figure}[]  	
	\vspace{0cm}
	\centering		  
	\includegraphics[width=1.0\columnwidth]{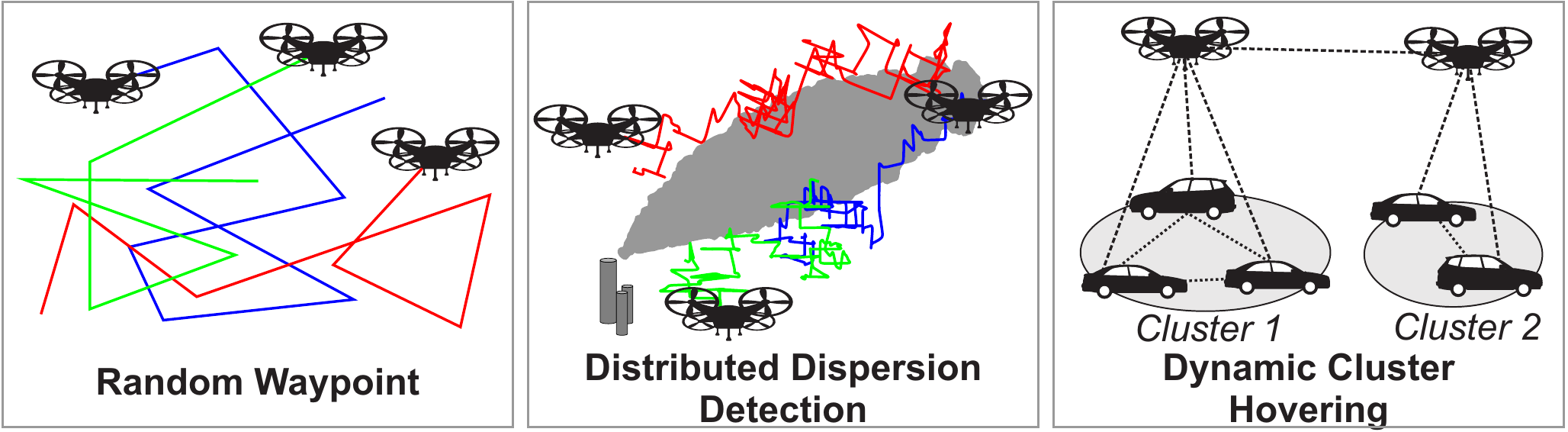}
	\vspace{-0.5cm}	
	\caption{Example trajectories and network topologies for the considered mobility algorithms.}
	\label{fig:mobility}
	\vspace{-0.5cm}	
\end{figure}
%
%
%

%
%
\textbf{Mobility models}: For analyzing the routing behavior of \protocol, a mixture of generic and realistic mobility models is considered. Visualizations of corresponding example trajectories and network topologies are shown in Fig.~\ref{fig:mobility}. 
\begin{itemize}
	%
	%
	\item \textbf{Random waypoint} is used an abstract reference scenario within the parameter selection process. In order to allow \protocol to exploit the trajectory knowledge and in contrast to conventional implementations, all future waypoint locations are computed at the beginning of the simulation run. 
	%
	\item \textbf{\ac{DDD}} \cite{Behnke/etal/2013a} is an algorithm for coordinated swarm-based plume source exploration in disaster scenarios which ensures intra-swarm connectivity through communication-aware mobility maneuvers.
	%
	%
	\item \textbf{Dynamic cluster hovering} is a method for \ac{UAV}-based network provisioning for hybrid vehicular networks initially presented in \cite{Sliwa/etal/2019c}. Hereby, multiple \acp{UAV} dynamically adjust their locations for providing network coverage for clusters of ground-based vehicles. Within the evalation scenario (which corresponds to the default scenario of the \ac{LIMoSim} mobility simulator \cite{Sliwa/etal/2019c}), a total number of $10$ \acp{UAV} operates at a flying height of 30~m. From a total number of 50 cars, a random subset of 10 vehicles is chosen to be equipped with communication interfaces. 
\end{itemize}
%
%
%

%
%
%
%
%
%
%
%
%
%

%
%
\textbf{Reference routing protocols}: In the next section, \protocol is compared to established routing protocols which implement different routing philosophies. All of the considered methods are configured according to their default parameter specification for mobile networking.
%
%
\begin{itemize}
	%
	%
	\item \textbf{\ac{AODV}} is a well-established \emph{reactive} routing protocol.
	%
	%
	\item \textbf{\ac{OLSR}} is a \emph{proactive} protocol which uses a path planning approach that involves information about the whole network topology. Moreover, the protocol utilizes so-called \acp{MPR} for minimizing the amount of broadcast messages within the routing message distribution process.
	%
	%
	\item \textbf{\ac{GPSR}} is \emph{geo-based} routing method which bases its greedy routing process on minimizing the geo-distance to the destination in each packet forwarding step.
	%
	%
	\item \textbf{B.A.T.Mobile} is a \emph{geo-predictive} extension to \ac{B.A.T.M.A.N.} and represents an immediate groundwork for \protocol.
\end{itemize}
According to the empirical analysis of \cite{Cavalcanti/etal/2018a}, the first three protocol represent the most commonly used approaches for \ac{VANET} routing. A summary about the simulation parameters is provided by Tab.~\ref{tab:parameters}.
%
%
\newcolumntype{L}{>{\raggedright\arraybackslash}X}
\newcolumntype{R}{>{\raggedleft\arraybackslash}X}

\begin{table}[ht]
	\centering
	\caption{Default Simulation Parameters}
	\vspace{-0.2cm}	
	\begin{tabularx}{\linewidth}{LR}
		\toprule
		\textbf{Parameter} & \textbf{Value} \\

		\midrule
	
		\ac{OMNeT++} version & 5.6.1 \\
		INETMANET version & 4.x \\
		\ac{MAC} & IEEE 802.11g \\
		Path loss exponent $\eta$ & 2.75 \\
		Channel model (Rural, Urban) & Friis, Nakagami (m=2) \\
		Scenario size (General) & 500~m $\times$ 500~m $\times$ 250~m   \\
		Scenario size (Cluster Hovering) & 750~m $\times$ 600~m $\times$ 50~m   \\
		
		Number of runs per configuration & 25 \\
		Simulation time & 900~s \\
		Number of routing hosts & 10~s \\
		Mobility update interval $\Delta t$ & 100~ms\\
		Prediction horizon $\tau$ & 2.5~s \\
		Waypoint radius $r_w$ & 10~m \\
		Velocity $v$ & 50~km/h \\
		
		Transport protocol & \acs{UDP} \\
		Traffic load per video stream & 2~MBit/s \\
		
		Learning rate $\alpha$ & 0.5 \\
		Basic discount factor $\gamma_0$ & 0.8 \\
		Chirp interval $\Delta t_{\text{chirp}}$ & 0.5~s \\
		
		\bottomrule
		
	\end{tabularx}
	\label{tab:parameters}
\end{table}
\vspace{-0.3cm}

\section{Simulation Results} \label{sec:results}

%
%
In this section, the results of the \ac{OMNeT++}-based simulations are presented. All errorbars show the 0.95 confidence interval of the mean value over the different simulation runs. In the following, the impact of different parameters on the behavior of \protocol is analyzed before the performance of the novel protocol is compared to established methods in defined reference scenarios.
%
%
%

%
%
\subsection{Parameterization of \protocol}

For all following parameter variations, the performance of \protocol is analyzed using the rural channel model and random waypoint mobility.

%
%
\begin{figure}[] 
	\centering
	\subfloat[]{\includegraphics[width=0.24\textwidth]{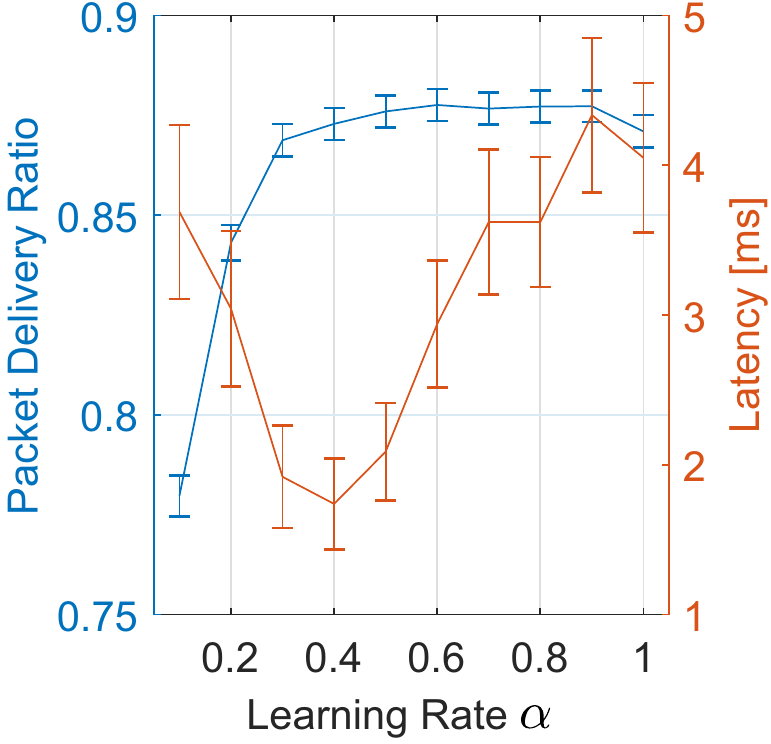}}\hfill
	\subfloat[]{\includegraphics[width=0.24\textwidth]{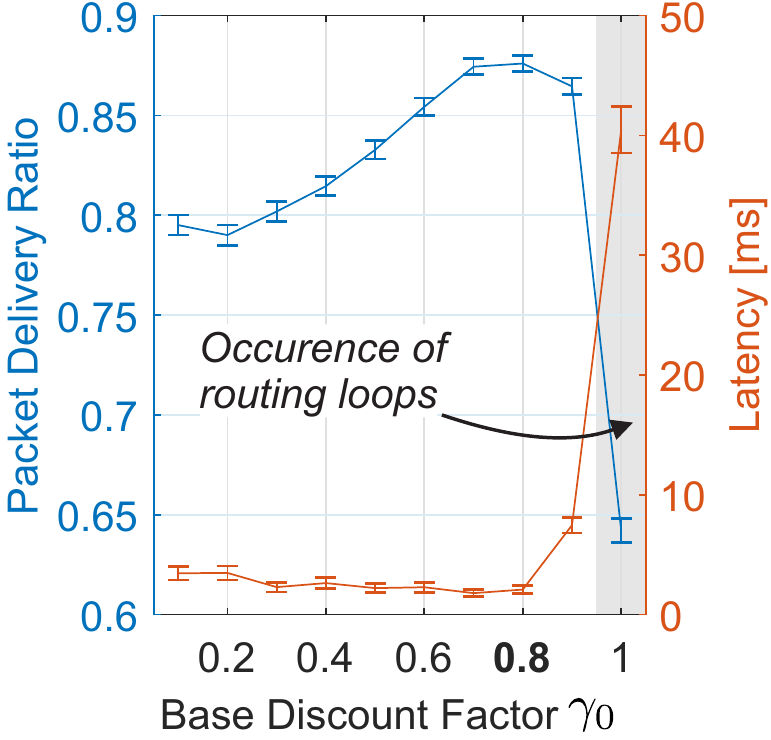}}\hfill		
	\caption{Impact of the reinforcement learning parameters on the end-to-end performance of \protocol.}
	\label{fig:learning_parameters}
	\vspace{-0.5cm}	
\end{figure}
%
%
\textbf{Learning parameters}: 
An evaluation of different values of the key reinforcement learning parameters of \protocol is shown in Fig.~\ref{fig:learning_parameters}.
%
%
The learning rate $\alpha$ corresponds to the information gain per received chirp message. Thus, it is dependent to the relative mobility of the agents and the chirp interval $\Delta t_{\text{chirp}}$. If $\alpha$ is chosen too small, the agent fails to adopt its decision making to the dynamics of the network topology. As a consequence of the resulting choice of sub-optimal routing paths, the end-to-end delay is increased and the \ac{PDR} is reduced. If $\alpha$ is chosen too large, the impact of single chirp messages --- which might be impacted by short-term effects such as local queuing  --- is overemphasized. In Fig.~\ref{fig:learning_parameters}~(a), it can be seen that $\alpha$ tolerates a certain grade of derivation from the optimal value without significantly reducing the end-to-end behavior of \protocol.

%
%
The basic discount factor $\gamma_0$, which is shown in Fig.~\ref{fig:learning_parameters}~(b), represents an implicit hop punishment for ensuring that the propagated reverse path quality to the originator decreases if the number of hops increases. Since the individual metrics that jointly form $\gamma(j)$ (see Eq.~\ref{eq:gamma}) are multiplied with each other, it also acts as a scaling factor for the latter. Similar to $\alpha$, $\gamma_0$ needs to be chosen large enough such that the information gain converges with the network dynamics. For $\gamma_0 = 1$, the loop free routing condition $\gamma(j)<1$ cannot be guaranteed. As a consequence, a massive drop of the end-to-end routing performance can be observed.
%
%
%

%
%
\textbf{Mobility prediction}:
%
%
\begin{figure}[]  	
	\vspace{0cm}
	\centering		  
	\includegraphics[width=1.0\columnwidth]{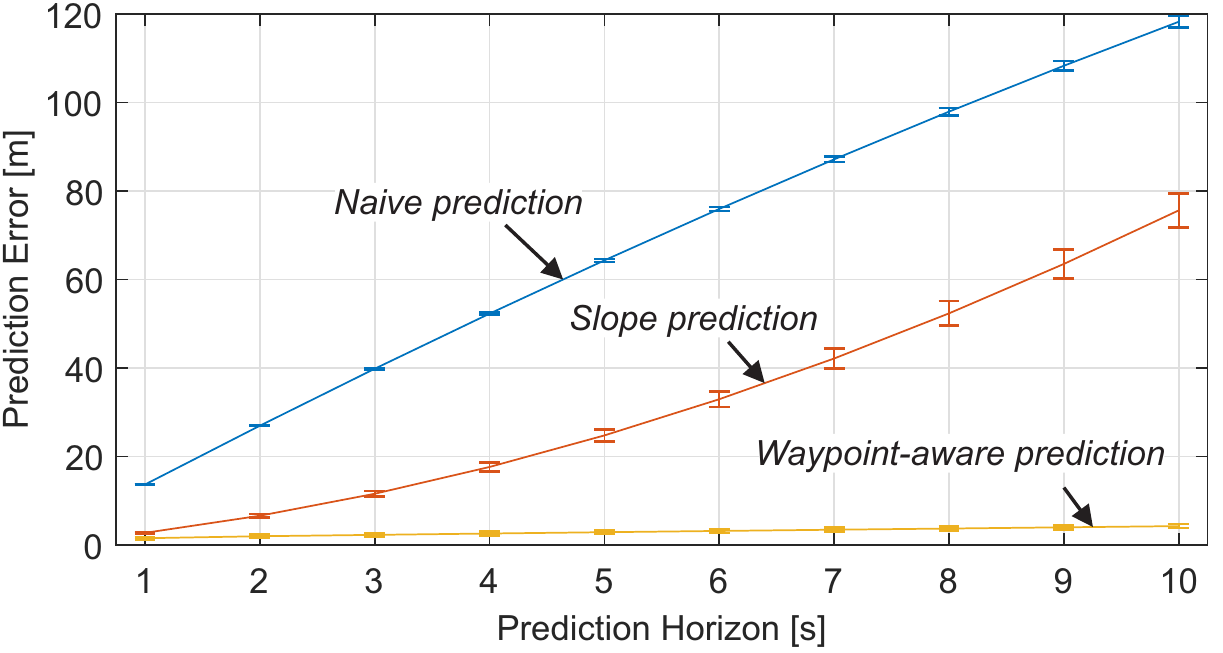}
	\vspace{-0.5cm}	
	\caption{Accuracy of different mobility prediction methods.}
	\label{fig:mobility_prediction_np}
	\vspace{-0.5cm}	
\end{figure}
For improving the communication robustness through consideration of the relative mobility of the agents, \protocol uses mobility prediction for deriving estimates of the corresponding \acp{LET}. Due to this dependency, the error of the mobility prediction should be minimized. Fig.~\ref{fig:mobility_prediction_np} shows the resulting errors for the considered mobility prediction schemes with respect to the prediction horizon $\tau$. 
%
%
As a reference, the behavior for a naive prediction method which assumes the vehicle position to stay constant is shown. In this case, the resulting error can be derived as $e_{\text{naive}} = v \cdot \tau$. In the worst case, the prediction points in the exact opposite direction of the actual movement which allows to derived an upper error bound $e_{\max} = 2 \cdot e_{\text{naive}}$.
%
%
It can be seen that the \emph{waypoint-aware} prediction allows to provide comparable accurate prediction results. In contrast to the extrapolation-based \emph{slope} method, it is able to integrate knowledge about the turning behavior of the vehicles into the prediction process.
%
%
For completeness, it is remarked that the integration of waypoints also strengthens the robustness of the protocol against inaccurate location information \cite{Sliwa/etal/2016b}.
As a consequence, \protocol is configured to prefer the waypoint-aware mobility prediction method. If the mobility model does not provide waypoint information, the slope prediction is utilized as a fallback.
%
%
%

%
%
\begin{figure}[] 
	\centering
	\subfloat[]{\includegraphics[width=0.24\textwidth]{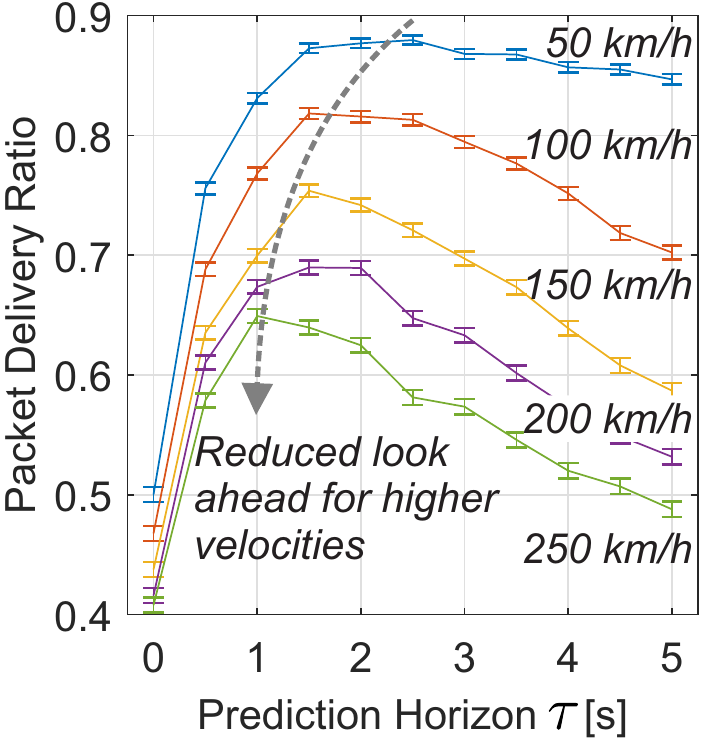}}\hfill
	\subfloat[]{\includegraphics[width=0.24\textwidth]{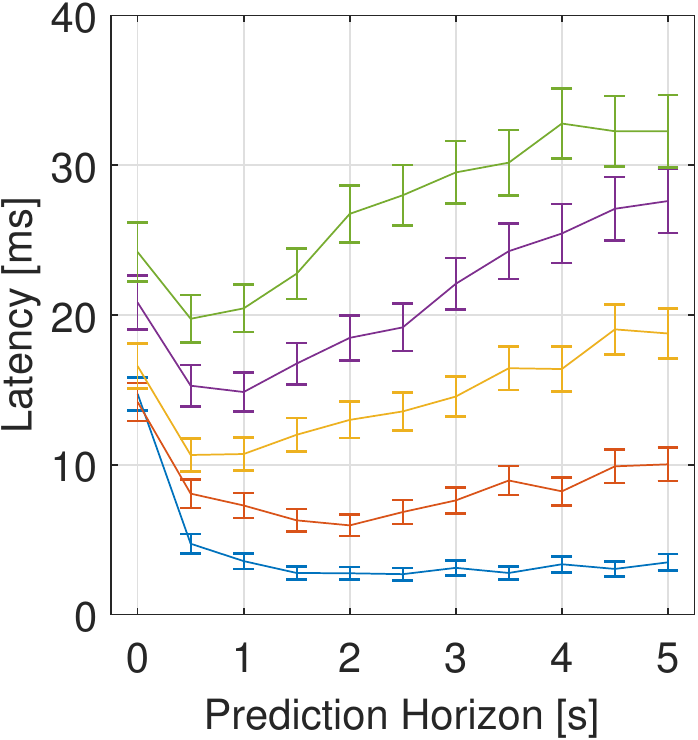}}\hfill		
	\caption{Impact of the prediction horizon $\tau$ on the behavior of \protocol at different velocities.}
	\label{fig:tau_speeds}
	\vspace{-0.5cm}	
\end{figure}
The impact of the prediction horizon $\tau$ on the behavior of \protocol for different velocities in the range of 50~km/h to 250~km/h is shown in Fig.~\ref{fig:tau_speeds}.
It can be seen that the link lifetime estimation of \protocol is highly depending on the availability of mobility prediction results. In comparison to the non-predictive variant ($\tau=0$), the \ac{PDR} is increased by up to 75~\% if the protocol uses a meaningful prediction horizon.
%
%
For higher speeds, the end-to-end routing performance decreases and smaller values of $\tau$ should be preferred in order to adopt to the increased network dynamics. In addition, \protocol becomes more sensitive to an optimal choice of $\tau$.
%
%
%

%
%
\subsection{Performance Comparison with Existing Routing Protocols}

%
%
\textbf{Random mobility}:
%
%
\begin{figure}[] 
	\centering
	\subfloat[]{\includegraphics[width=0.24\textwidth]{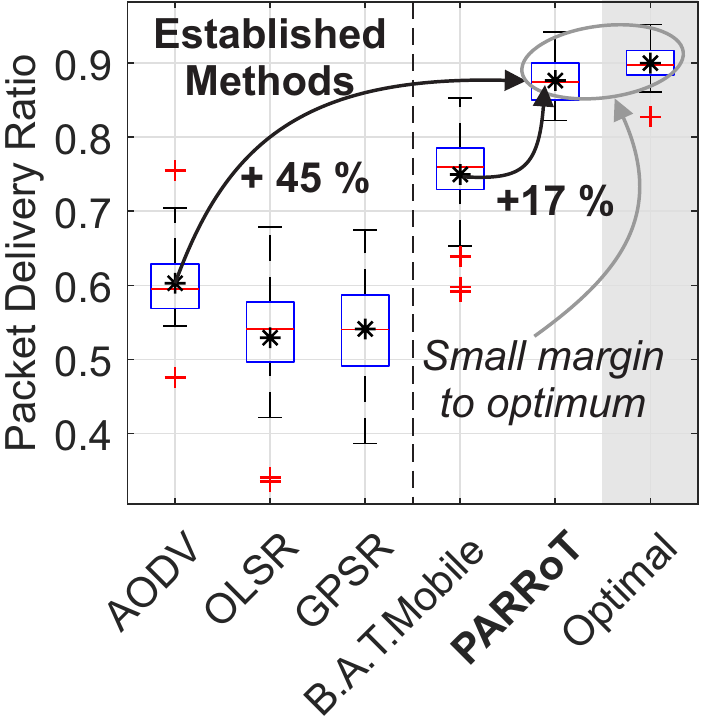}}\hfill
	\subfloat[]{\includegraphics[width=0.24\textwidth]{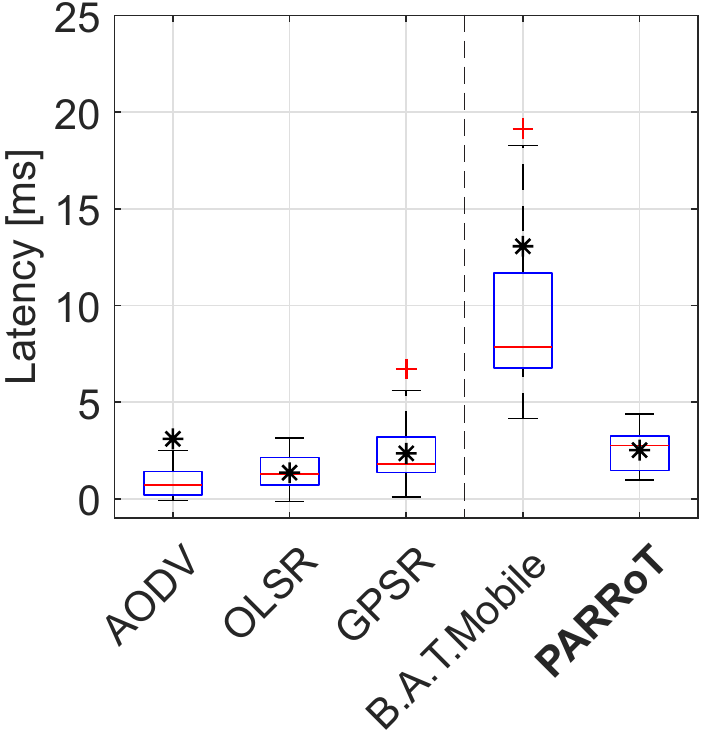}}\hfill		
	\caption{Comparison \protocol and existing routing methods. \emph{Optimal} represents an upper topology-related bound for the path availability between sender and destination (Rural radio propagation model).}
	\label{fig:random_waypoint_friis}
	\vspace{-0.5cm}
\end{figure}
Fig.~\ref{fig:random_waypoint_friis} show the performance of the considered routing methods using random waypoint mobility with the rural channel model. 
%
%
Due to the non-coordinated motion, the existence of a routing path between sender and receiver is not always guaranteed. In order to pay attention to this aspect while classifying the performance of the different protocols, a theoretical upper bound --- referred to as \emph{Optimal} --- for the \ac{PDR} is provided based on post processing-based analysis of the network topology. However, it is remarked that this method is only able to consider the mobility-related aspects and does not account for load-related packet loss.
In comparison to the well-established methods \ac{AODV}, \ac{OLSR}, and \ac{GPSR}, \protocol show a massively higher --- at least by 45~\% --- \ac{PDR} which is close to the optimum robustness. Moreover, its reinforcement learning-based routing approach even outperforms the also prediction-based routing method \batmobile by 17~\%.
%
%
%

%
%
\begin{figure}[] 
	\centering
	\subfloat[]{\includegraphics[width=0.24\textwidth]{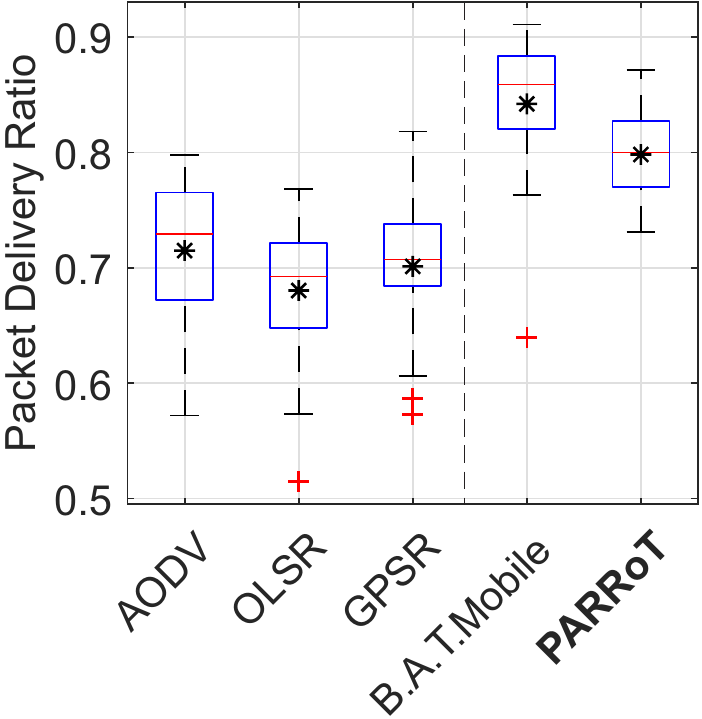}}\hfill
	\subfloat[]{\includegraphics[width=0.24\textwidth]{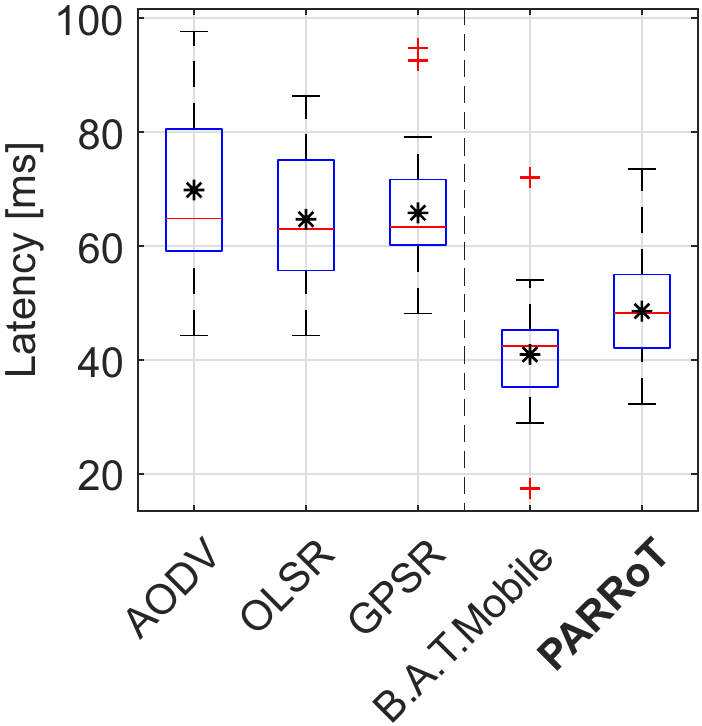}}\hfill		
	\caption{Comparison \protocol and existing routing methods (Urban radio propagation model).}
	\label{fig:random_waypoint_nakagami}
	\vspace{-0.7cm}
\end{figure}
More challenging radio channel conditions are analyzed in Fig.~\ref{fig:random_waypoint_nakagami}. As the utilized channel model is subject to probabilistic effects, an upper bound for the \ac{PDR} cannot be derived.
%
%
Here, \protocol shows a slightly lower \ac{PDR} than \batmobile. A plausible explanation for this observation is that the fast fading effects reduce the significance of the neighbor set coherence metric which then leads to sub-optimal routing decisions. Future extensions of \protocol could explicitly address this issue through dynamic channel-dependent weighting the different metrics.
%
%
All protocols suffer from a higher end-to-end delay due to an increased queuing time at the \ac{MAC} layer due to sporadic link loss. The general tendency of these findings is also confirmed by the experiments of \cite{Saleh/etal/2020a}.
%
%
%

%
%
\begin{figure}[]  	
	\vspace{0cm}
	\centering		  
	\includegraphics[width=1.0\columnwidth]{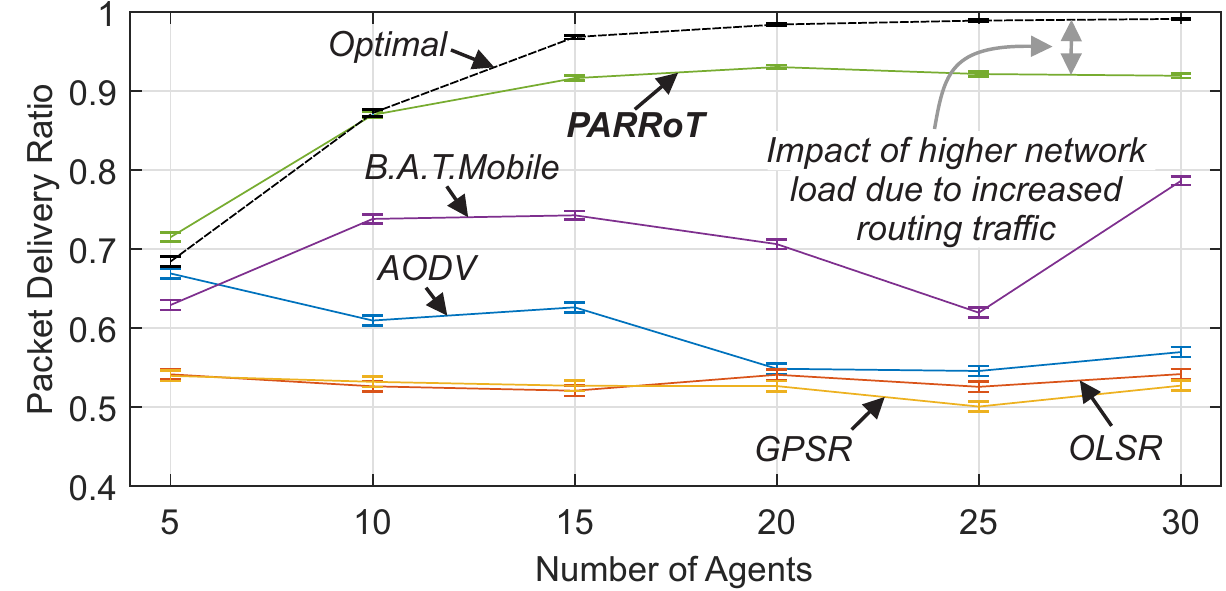}
	\vspace{-0.7cm}	
	\caption{Scalability analysis: Impact of the number of hosts on the end-to-end routing performance.}
	\label{fig:num_hosts}
	\vspace{-0.7cm}	
\end{figure}
A scalability analysis of the considered routing protocols is shown in Fig.~\ref{fig:num_hosts}.
%
%
As the size of the scenario is not varied, the increase of the number of agents corresponds to increasing the density of network nodes and the grade of routing opportunities. However, the higher amount of periodic routing traffic --- especially for the proactive methods but also for neighbor discovery in reactive routing --- leads to an increased probability for collision-related packet loss.
%
%
It can be seen that the established routing methods fail to adopt to the network topology dynamics. Due to the high dependency of individual routing messages, they are unable to exploit the higher amount of possible routing paths. 
%
%
Due to the inherent maintenance of different routing paths per destination, \protocol is robust against loss of individual chirp messages. In contrast to the reference protocols, it is able to leverage the increased network density for improving the robustness of the data transfer. However, for more than 20 active hosts, a slight decrease of the \ac{PDR} can be observed.
%
%
%

%
%
\textbf{Application-driven mobility}:
%
%
\begin{figure}[] 
	\centering
	\subfloat[\ac{DDD}-based Swarm Exploration]{\includegraphics[width=0.24\textwidth]{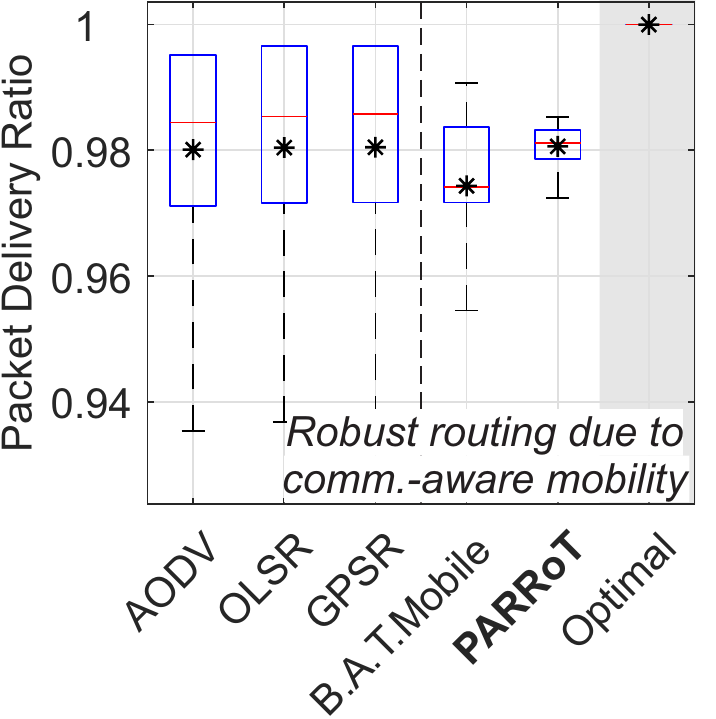}}\hfill
	\subfloat[Cluster Hovering]{\includegraphics[width=0.24\textwidth]{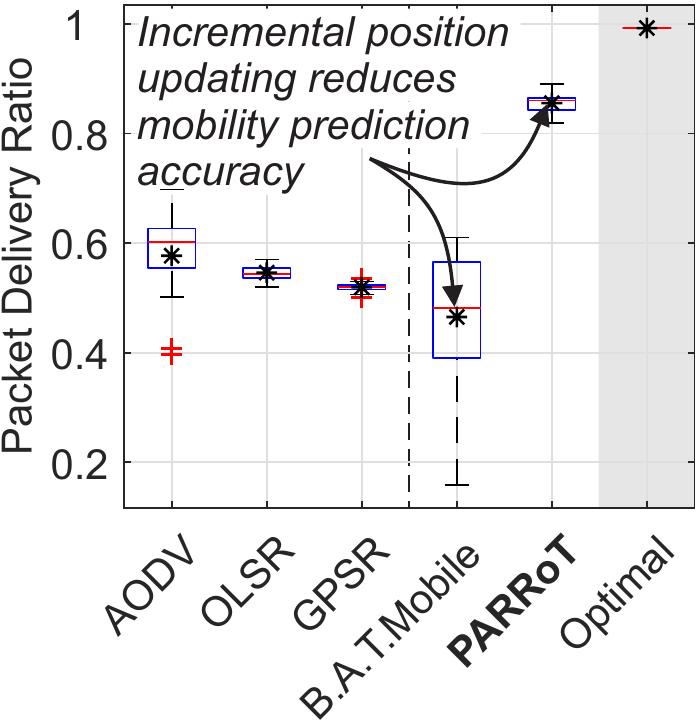}}\hfill		
	\caption{Routing robustness in realistic mobility conditions (Rural channel model).}
	\label{fig:realistic_mobilty}
	\vspace{-0.7cm}
\end{figure}
The behavior of the protocols in more realistic and application-specific mobility conditions is shown in
Fig.~\ref{fig:realistic_mobilty}.
%
%
For the swarm mobility exploration method \ac{DDD}, the corresponding results are visualized in Fig.~\ref{fig:realistic_mobilty}~(a). Due to implemented \emph{communication-aware} mobility approach which proactively adjusts the mobility behavior of the agents for ensuring swarm coherence, the \ac{PDR} shows a high general level and does not vary significantly between the different routing protocols.
Still, the predictive approach of \protocol allows to proactively avoid some of the lower outlier values.
%
%
%

%
%
The results for the \ac{UAV}-based cluster hovering analysis are shown in Fig.~\ref{fig:realistic_mobilty}~(b).
%
%
This evaluation scenario is characterized by a complex network topology with a large number of mobile vehicles which show different mobility characteristics. In addition, the \ac{UAV} cluster hovering is based on a ground traffic-related incremental position updating process which does not allow to accurately forecast future locations. As a consequence, the mobility-predictive routing methods are confronted with imprecise estimations for the relative mobilities.
%
%
In this complex setting, \batmobile is not able to maintain robust connectivity. In fact, the mobility predictions even reduce the routing performance of the protocol slightly beyond the \ac{PDR} of the established methods. For completeness, it is remarked that these issues might be partially compensable through a scenario-specific parameter optimization.
%
%
Although, \protocol is also confronted with the same challenges, the resulting degradation of the routing performance is far less distinct. Due to the implemented multi-metric routing approach which does not only consider the link lifetime but also the neighbor set coherence, it is less vulnerable to imprecise mobility predictions. As a consequence, \parrot is able to provide robust data delivery even in highly challenging hybrid vehicular networks.

\section{Conclusion}

%
%
In this paper, we presented the novel routing protocol \protocol for highly mobile robotic networks which brings together mobility-predictive routing with reinforcement learning-based decision making. 
%
%
In a comprehensive simulation-based performance evaluation, it was shown that the consideration of the future relative mobility between the agents allows \protocol to achieve robust and efficient data delivery even in challenging radio propagation conditions.
%
%
%

%
%
In future work, we will extend \protocol for considering environment information within the link lifetime estimation process by integrating knowledge about the surrounding obstacles such as buildings. In addition, we will focus on analyzing the real world performance of the novel routing protocol.

\ifdoubleblind

\else

	\section*{Acknowledgment}
	
	\footnotesize
	This work has been supported by the German Research Foundation (DFG) within the Collaborative Research Center SFB 876 ``Providing Information by Resource-Constrained Analysis'', projects A4 and B4 as well as by the German Federal Ministry of Education and Research (BMBF) in the project A-DRZ (13N14857).

\fi

\ifacm
\else
	\bibliographystyle{IEEEtran}
	\bibliography{Bibliography}
\fi

\end{document}